\documentclass[12pt]{article}
\usepackage{amssymb,epsf,cite}
\makeatletter
\if@titlepage
  \renewcommand\maketitle{\begin{titlepage}%
  \let\footnotesize\small
  \let\footnoterule\relax
  \null\vfil
  \vskip 60\p@
    \begin{center}%
    {\Large \bfseries\@title \par}%
    \vskip 3em%
    {\normalsize
     \lineskip .75em%
      \begin{tabular}[t]{c}%
        \@author
      \end{tabular}\par}%
      \vskip 1.5em%
    {\normalsize \@date \par}
  \end{center}\par
  \@thanks
  \vfil\null
  \end{titlepage}%
  \setcounter{footnote}{0}%
  \let\thanks\relax\let\maketitle\relax
  \gdef\@thanks{}\gdef\@author{}\gdef\@title{}}
\else
\renewcommand\maketitle{\par
  \begingroup
    \renewcommand\thefootnote{\fnsymbol{footnote}}%
    \def\@makefnmark{\hbox to\z@{$\m@th^{\@thefnmark}$\hss}}%
    \long\def\@makefntext##1{\parindent 1em\noindent
            \hbox to1.8em{\hss$\m@th^{\@thefnmark}$}##1}%
    \if@twocolumn
      \ifnum \col@number=\@ne
        \@maketitle
      \else
        \twocolumn[\@maketitle]%
      \fi
    \else
      \newpage
      \global\@topnum\z@   
      \@maketitle
    \fi
    \thispagestyle{plain}\@thanks
  \endgroup
  \setcounter{footnote}{0}%
  \let\thanks\relax
  \let\maketitle\relax\let\@maketitle\relax
  \gdef\@thanks{}\gdef\@author{}\gdef\@title{}}
\def\@maketitle{%
  \newpage
  \null
  \vskip 1em%
  \begin{center}%
    {\Large \bfseries\@title \par}%
    \vskip 1.5em%
    {\normalsize
      \lineskip .5em%
      \begin{tabular}[t]{c}%
        \@author
      \end{tabular}\par}%
    \vskip 1em%
    {\normalsize \@date}%
  \end{center}%
  \par
  \vskip 1.5em}
\fi
\renewcommand\section{\@startsection{section}{1}{\z@}%
                                     {-3.25ex\@plus -1ex \@minus -.2ex}%
                                     {1.5ex \@plus .2ex}%
                                     {\reset@font\normalsize\bfseries}}
\renewcommand\subsection{\@startsection{subsection}{2}{\z@}%
                                    {3.25ex \@plus1ex \@minus.2ex}%
                                    {-1em}%
                                    {\reset@font\normalsize\bfseries}}
\@addtoreset{equation}{section}
\newcommand{\note}[1]{\raisebox{1ex}{{\footnotesize \sf #1}}}
\newcommand{\rnote}[1]{\raisebox{1ex}{{\hspace*{-3mm} \scriptsize\sf#1}}
                       \hspace*{-4mm}}
\makeatother
\def\be{ \begin{equation}}          \def\ee{ \end{equation}}
\def\ba{ \begin{eqnarray}}          \def\ea{ \end{eqnarray}}
\renewcommand{\theequation}{{\thesection.\arabic{equation}}}
\def\nn{\nonumber}                  

\def\C{\mathbb{C}}  \def\R{\mathbb{R}}\def\T{\mathbb{T}}
\def\o{\otimes}                     

\def\cedille#1{\setbox0=\hbox{#1}\ifdim\ht0=1ex \accent'30 #1%
 \else{\ooalign{\hidewidth\char'30\hidewidth\crcr\unbox0}}\fi}
\def\gaw{Gaw\cedille edzki}

\def\Ad{\mbox{\rm Ad}} \def\ad{\mbox{\rm ad}} 

\def\mathR{\R}

\def\a{\alpha }          \def\b{\beta }

\def\ew{\hspace*{-1mm}}   \def\ppe{\hspace*{-2.5mm}}

\newcommand{\Fus}[6]{F_{{\scriptstyle #1}{\scriptstyle #2}}
  \hspace*{.3mm}\displaystyle{[} \ew \begin{array}{ll} {\scriptstyle #3 }
  \ppe & {\scriptstyle #4} \ppe \\[-2mm] {\scriptstyle #5}\ppe &
  {\scriptstyle #6}\ew \end{array}\displaystyle{]}}
\newcommand{\CG}[6]{\displaystyle{[} \,\ew \begin{array}{lll} 
  {\scriptstyle #1} \ppe
  & {\scriptstyle #2} \ppe & {\scriptstyle #3} \ew \\[-2mm] {\scriptstyle
  #4} \ppe & {\scriptstyle #5}\ppe & {\scriptstyle #6} \ew\end{array}
  \displaystyle{]}}
\newcommand{\SJS}[6]{ \displaystyle{\{ } \ew \begin{array}{lll} 
  {\scriptstyle #1} \ppe  &
  {\scriptstyle #2} \ppe & {\scriptstyle #3}
  \ppe \\[-2mm]{\scriptstyle #4}  \ppe & {\scriptstyle #5} \ppe &
 {\scriptstyle #6} \ew \end{array} \displaystyle{\} } }

\def\ik{{\sf k}}
\def\min{{\mbox{\rm min\/}}}
\def\cH{{\cal H}}

\def\nno{\newline\noindent}
\def\id{{\rm id}}
\title{Non-commutative World-volume Geometries:\\[2mm] 
Branes  on   SU(2) and Fuzzy Spheres}
\author{{\sc Anton Yu.\ Alekseev$\,$ \rnote{1},  
           \  Andreas Recknagel$\,$ \rnote{2},}\\[1mm] 
         {\sc   Volker Schomerus 
\rnote{3} } \\[9mm] 
\note{1} Institute for Theoretical Physics, Uppsala University  
\\ Box 803, S--75108 Uppsala, Sweden
\\[1mm]
\note{2} Max-Planck-Institut f\"ur Gravitationsphysik, 
       Albert-Einstein-Institut 
\\ Am M\"uhlenberg 1, D--14424 Potsdam, Germany
\\[1mm]
\note{3} II. Institut f\"ur Theoretische Physik, Universit\"at Hamburg
\\ Luruper Chaussee 149, D--22761 Hamburg, Germany
}
\date{July 31, 1999}
\begin{document}
\begin{titlepage}      \maketitle       \thispagestyle{empty}

\vskip1cm
\begin{abstract}
\noindent  
The geometry of D-branes can be probed by open string 
scattering. If the background carries a non-vanishing 
B-field, the world-volume becomes non-commutative. 
Here we explore the quantization of world-volume
geometries in a curved background with non-zero 
Neveu-Schwarz 3-form field strength $H = dB$. Using exact 
and generally applicable methods from boundary conformal 
field theory, we study the example of open strings in 
the SU(2) Wess-Zumino-Witten model, and establish a 
relation with fuzzy spheres or certain (non-associative) 
deformations thereof. These findings could be of direct 
relevance for D-branes in the presence of Neveu-Schwarz 
5-branes; more importantly, they provide insight into a 
completely new class of world-volume geometries. 
\end{abstract}
\vspace*{-20.9cm}
{\tt {DESY 99-104 \hfill AEI 1999-11}}\\
{\tt {ESI 755 (1999) \hfill hep-th/9908040}}\break
\bigskip\vfill
\noindent\phantom{wwwx}{\small e-mail: }{\small\tt alekseev@teorfys.uu.se, 
anderl@aei-potsdam.mpg.de,}\\  
\phantom{wwwx{\small e-mail: }}{\small\tt vschomer@x4u.desy.de} 
\end{titlepage}
\section{Introduction} 

It was observed by Douglas and Hull \cite{DoHu} that D-branes on $\T^2$ 
with a constant Neveu-Schwarz (NS) two-form potential $B$ give rise to 
an effective world-volume theory on a non-commutative torus. Even though 
this initial observation was re-considered and generalized by many 
authors \cite{NCTor,ChHo,Vol}, all the subsequent work is restricted 
to flat backgrounds. A perturbative analysis along the lines of 
\cite{Vol}, on the other hand, shows that the quantization of 
world-volume geometries should be a much more general phenomenon 
which persists in the case of curved backgrounds. 

\smallskip\noindent
In this work we shall present the first non-perturbative (in $\alpha'$) 
investigation of world-volume geometries in a curved string background 
with non-vanishing NS 3-form field $H = dB$. \footnote{Recall that the 
curvature is linked to the field strength $H$ by the string's equation 
of motion.} 
An exact treatment of D-branes in curved backgrounds is possible within  
the framework of boundary conformal field theory. Here we illustrate the
basic techniques and some general features of the resulting world-volume 
geometries in a particular example, namely the SU(2) WZW theory, 
and study D-branes in the WZW model associated with the gluing condition 
$J^a = \bar J^a$. We shall argue that their world-volumes may be regarded 
as fuzzy two-spheres when the level $\ik$ is sent to infinity, i.e.\ when 
the background becomes flat. For finite level, $H$ is non-zero and we shall 
find non-associative deformations of these fuzzy spheres, which are closely 
linked to the theory of quantum groups. While the infinite level result can 
be predicted from the semi-classical analysis in \cite{AlSc} together with 
the general phenomenon of world-volume quantization in flat backgrounds 
\cite{DoHu}, our results on the finite level provide a non-trivial 
extension of the standard rules. Apparently, many features of the world-%
volume geometry are not captured by the perturbative treatment of D-branes 
on group manifolds that was suggested recently in \cite{CoPl}. 

\smallskip\noindent
We shall follow a general procedure  which allows us to extract 
world-volume geometry from the world-sheet description of any (generalized) 
D-brane, even when it is given in purely algebraic terms. The essential 
input data are the operator product expansions (OPE) of boundary fields 
(open string vertex operators). Since they depend on the ordering of 
the operators, it is not surprising that the brane world-volume obtained 
in this way is a non-commutative space, in general. We shall see that 
non-associativity may show up as well. 

\begin{figure} 
{\hspace{3cm}\epsfbox{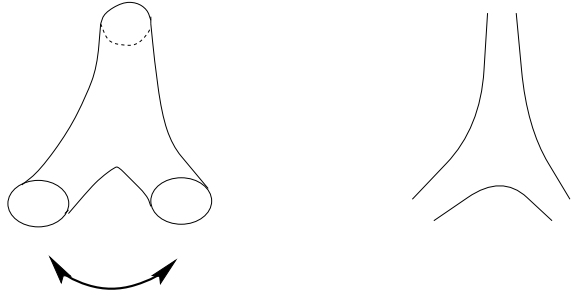}} \vspace*{2mm}
\caption{ \small 
World-sheet diagrams for closed resp.\ open string interaction.
Having assigned vertex operators to the legs, they can be read as 
structure constants for the multiplication of two operators, projected on 
the third channel. In the closed string case, the in-coming operators 
can be interchanged with the help of world-sheet diffeomorphisms, while 
the ordering of open string vertices is fixed up to cyclic permutations.} 
\end{figure}  
\medskip\noindent 
Our approach is inspired by a project initiated by J.\ Fr\"ohlich 
and K.\ \gaw\ in \cite{FG} (see also \cite{CF} for earlier ideas in 
the same direction), where it was proposed to construct non-commutative 
target space geometries from OPEs of closed string vertex operators. 
This was developed further in \cite{FGR,Gr}. It appears, however, 
that non-commutative geometry emerges in a more natural way and 
on a more fundamental level in the open string case, cf.\ the picture below. 
\noindent 
Our findings add to the growing evidence that brane physics surpasses 
classical geometry -- even though the 
emergence of a non-commutative world-volume need not necessarily mean 
that a D-brane behaves non-geometrically in the sense of the criterion 
formulated in \cite{BDLR}. This criterion rests on a comparison of  
low-energy effective field theories in the stringy and in the
large-volume regime, and we do not attempt to test it in the present 
paper. But we would like to point out that the structures contained in 
the non-commutative world-volume also form the main ingredient of the 
effective action of the brane. 

\noindent 
While we have chosen the SU(2)$_\ik$ example mainly because of its 
simplicity and because there exists a semi-classical curved background 
picture, it is also an important ingredient of the CFT 
formulation of the Neveu-Schwarz 5-brane, see e.g.\ \cite{fivebr}.  
Given that questions like stability of the configuration 
can be clarified, our findings should be relevant for the geometry 
of D-branes in the presence of a stack of 5-branes. Similarly, our 
SU(2) WZW results could be applicable in the study of branes on an 
AdS$_3 \times S^3$ string background, see e.g.\ \cite{GKS,Te}. 

\medskip

\section{World-volume geometry -- from the flat case to arbitrary 
         backgrounds} 

Before we show how one can read off fuzzy geometry from branes 
in the WZW model, let us briefly review the emergence of non-commutative 
spaces in the more standard case of branes in flat $n$-dimensional 
Euclidean space $\mathR^n$, or on a flat torus $\T^n$. Consider a 
D-brane which is localized along a $p$-dimensional hyper-plane $V_p$ 
in the target, with tangent space $TV_p$.  The conformal field theory 
associated with such a Euclidean D-brane is defined on the upper half 
of the complex plane. It contains an $n$-component free bosonic field 
$X= (X^\mu(z,\bar z)),\ \mu = 1,\dots,n,$ subject to Neumann boundary 
conditions in the directions along $TV_p$ and Dirichlet 
boundary conditions for components perpendicular to the world-volume 
of the brane. From the free bosons, one may obtain various new fields, 
in particular the open string vertex operators  
$$ 
V_k (x) = \ :\exp(i k X(x)) :  \ \quad \mbox{ for all } \ \ \ 
    k \in TV_p \ \ ,
$$
which can be inserted at any point $x$ on the real line. When there is 
no magnetic field on the brane, the OPE of these U(1)-primaries reads  
(with $\alpha' = \frac12$ and for $x_1 > x_2$) 
\be V_{k_1}(x_1)\, V_{k_2} (x_2) \ =\ (x_1-x_2)^{k_1k_2/2} \ 
        V_{k_1+ k_2}(x_2)  + \dots \ \ , \label{OPEB0} \ee
where the dots indicate less singular non-primary contributions. 
We can rewrite this relation by introducing the objects    
$$ 
f(X(x)) \equiv V[f](x) \ : = \ \frac{1}{(2\pi)^{p/2}} 
    \int_{TV_p} d^pk \ \hat f(k) \ V_k(x) 
$$
for each function $f : V_p \rightarrow \C$ with  Fourier transform
$\hat f(k)$. Then the boundary OPE (\ref{OPEB0}) translates into a 
``definition'' of pointwise multiplication of functions,  
\be
V[\,f\,](1)\; V[\,g\,](0) = V[\,f\cdot g\,] (0) + \ldots \quad . 
\label{pointw}\ee
We have specialized to coordinates $x_1=1$ and $x_2=0$ for convenience, 
arbitrary insertion points can be recovered via conformal covariance.  

\noindent
The effect of switching on a $B$-field is described by adding the term 
\be S_B \ = \ \frac{1}{2\pi} \ \int dz d\bar z\, B_{\mu\nu}\partial 
    X^\mu(z,\bar z) \bar\partial X^\nu(z,\bar z) \label{Bterm}\ee
to the action of the original theory without $B$-field. One can 
easily see that this is a pure boundary term with no influence
on the bulk properties of the theory. It only changes the boundary 
conditions. If we assume for definiteness that $V_p$ is 
spanned by the first $p$ coordinates $x^\mu,\ \mu = 1, \dots, p$,
the new boundary conditions read (with $ z = x + i y$)
\be \partial_y X^\mu (z,\bar z) \ = \ B^{\mu}_{\ \nu} 
        \partial_x X^\nu(z, \bar z)   \ \ \ \mbox{ for } \ \ \ \ 
     z = \bar z  \ \ \mbox{ and } \ \ \ \mu,\nu = 1, \dots,p\ \ . 
\label{Bglue} \ee   
This means that the (exact) free boson propagator becomes ($x_1,x_2
\in\mathR$) 
\be
\langle \, X^\mu(x_1)X^\nu(x_2)\,\rangle_{B} 
= - \,\bigl(\delta^{\mu\nu} + \Theta_S^{\mu\nu}\bigr)\; 
 \log|x_1-x_2| - i\,\frac{\pi}{2}  \, \Theta_{\! A}^{\mu\nu}\; 
   {\rm sign}(x_1-x_2) \label{Xcorr}\ee
where $ \Theta_S$ and $\Theta_A$ denote the symmetric resp.\ 
anti-symmetric part of the matrix 
$\Theta = (1-B)(1+B)^{-1}\,$. Explicitly,
\be
\Theta_A= \frac{2}{B-B^{-1}}\ .
\label{magic}\ee
In particular, when $B$ is large we obtain $\Theta_A \approx
2 B^{-1}$, which means that $\Theta_A$ is the Poisson
bi-vector corresponding to the symplectic form $B$.
Eq.\ (\ref{Xcorr}) immediately yields the boundary OPE for a 
non-vanishing $B$-field, 
$$ 
V_{k_1}(1) \, V_{k_2}(0) \ = \ e^{- i \frac{\pi}{2}\; k_1^{\rm t} 
  \Theta_A  k_2} \;V_{k_1+k_2}(0) + \ldots\ \ .
$$
As before, this can be used to define a (deformed) product $\star$ for 
functions through $V[f](1) V[g](0) = V[f \star g] (0) + \ldots\; ,$
where now  
\be
\bigl(\,f \star g \,\bigr) (x) :=  e^{i\,\frac{\pi}{2}\;
\Theta^{\mu\nu}_A 
\partial^x_\mu\partial^y_\nu}\; f(x) g(y)\; \big\vert_{y=x} \ \ . 
\label{moyal}\ee
This is the associative, non-commutative Moyal-Weyl product of 
functions $f,g$ on the world-volume $V_p$ of the brane.  In the 
context of the derivation we have given, non-commutativity of $\star$ 
arises because the ordering of boundary fields in general does 
matter, cf.\ the sign-term in eq.\ (\ref{Xcorr}). The algebra 
of functions with product (\ref{moyal}) is, of course, the 
non-commutative brane world-volume uncovered by Douglas and 
Hull using a different approach.  It is a deformation of the 
ordinary algebra of functions, with deformation parameter(s) 
given by (the matrix) $\Theta_A$. 

\medskip\noindent 
In \cite{Vol}, the term (\ref{Bterm}) was viewed as a bulk perturbation of 
the $B=0$ theory, i.e.\ techniques of conformal perturbation theory 
were applied to the operator $\exp(-S_B)$ being inserted into arbitrary 
correlation functions of the $B=0$ theory. This perturbative analysis, 
which can be extended to arbitrary $\sigma$-models (at least in the 
case $dB=0$), leads to a string theoretic picture of Kontsevich's 
quantization of Poisson manifolds \cite{Kon}, see also the work of 
Cattaneo and Felder \cite{CatFel}. It clearly displays that the 
quantization of world-volume geometries should be expected beyond 
the case of constant $B$-fields. This will be confirmed through our exact 
analysis of the WZW model (see discussion of the limit $\ik \rightarrow 
\infty$ below). As we remarked in the introduction, new phenomena are 
bound to occur when $dB$ does not vanish. In such cases, the classical 
world-volume of a brane comes equipped with some generalization of an
ordinary 
Poisson-structure, and there exists no general notion of ``quantization'' 
for such geometries. Hence, the investigation of 
branes in a non-vanishing NS 3-form field strength $H = dB$ can teach 
us new lessons on how to quantize certain non-Poisson geometries. In 
our example of branes on SU(2) we shall recover some variants of 
well-known quantum group algebras.   

\medskip\noindent 
Our formulation of the simple example of flat branes in a constant 
$B$-field motivates the following {\sl general procedure}: 
When we want to associate 
non-commutative spaces to branes which are given as boundary conditions 
on the world-sheet, we take the OPE of boundary fields 
(open string vertex operators corresponding to internal excitations 
of the brane) as a basic input. Then we choose a  
suitable subset of boundary fields (e.g.\ primaries as above) and 
use them as abstract generators of an algebra of ``functions'' 
on the (non-commutative) world-volume of the brane, with multiplication 
table given by the boundary OPE (projected onto the subset, and 
evaluated at $x_1=1$ and $x_2=0$, say). 

\noindent
Further comments on this general prescription will be given later, but 
now we would like to test it in the case of SU(2) WZW models, where 
the semi-classical picture provides certain expectations as to how 
the ``quantized world-volume'' of branes should look like.

\section{D-branes in the SU(2) WZW model}

\subsection{Semi-classical analysis.}
The SU(2) WZW model at level $\ik$ describes strings moving on a 
three-sphere $S^3$ of radius $R \sim \sqrt{\ik}$, which is equipped 
with a constant NS 3-form field strength 
$$
H \ \sim\ \frac{1}{\sqrt{\ik}}\;  \Omega \ =\
 \frac{1}{\sqrt{\ik}}\;  f_{abc}\; \theta^a \wedge \theta^b \wedge \theta^c \ ,
$$
where $\Omega$ denotes the usual volume form on the unit 
sphere, and $\theta^a$ are components of the 1-form $dgg^{-1}$.
 In superstring theory, this geometry appears in the space 
transverse to a stack of $\ik$ NS 5-branes. These branes act as 
sources for $\ik$ units of NS 3-form flux through a three-sphere 
surrounding their (5+1)-dimensional world-volume.  

\noindent 
The world-sheet swept out by an open string in $S^3$ is 
parametrized by a map $g: {\rm H} \rightarrow {\rm SU(2)}$ from the 
upper half-plane H into the group manifold SU(2)$\, \cong S^3$. From 
this field $g$ one obtains Lie algebra valued chiral currents 
$$ J(z) \ = \ - \ik \, (\partial g) g^{-1} \ \ \ , \ \ \ \
   \bar J(\bar z) \ = \ \ik \, g^{-1} \bar\partial g \ \  $$
as usual. We shall be interested in maximally symmetric D-branes on 
SU(2), which are characterized by the gluing condition $J(z ) = \bar J 
(\bar z)$ along the boundary $z = \bar z$. They were analyzed 
from a semi-classical point of view in \cite{AlSc}, and we 
shall briefly recall the findings of this approach. (For a detailed 
path integral description of branes in SU(2), see \cite{Gaw}.)
\nno 
We first decompose the tangent space $T_h{\rm SU}(2)$ at 
each point $h \in \,$SU(2) into a part $T^{||}_h{\rm SU}(2)$ tangential to 
the conjugacy class through $h$ and its orthogonal complement 
$T^\perp_h {\rm SU}(2)$ (with respect to the Killing form). 
In \cite{AlSc}, the following two basic observations were made:
\begin{enumerate} 
\item With gluing conditions of the type $J = \bar J$, the endpoints
 of open strings on SU(2) are confined to conjugacy 
 classes, i.e.\ 
$$ (g^{-1} \partial_x g)^\perp \ = \ 0 \ \ .$$
\item Along the individual branes, i.e.\ along the conjugacy classes of 
SU(2), the gluing condition becomes
$$ (g^{-1} \partial_y g)^{||} \ = \ \frac{\Ad(g)+1}{\Ad(g)-1} 
   (g^{-1} \partial_x g)^{||} \ \ . $$ 
\end{enumerate} 
Except for two degenerate cases, namely the points $e$ and $-e$ on 
the group manifold, the conjugacy classes are two-spheres in SU(2). 
Taking into account the usual correspondence between 
$\sqrt{\ik} g^{-1} \partial g$ and the flat space coordinate
$\partial X$, recalling that the metric on the three-sphere scales with 
$\ik$, and comparing with the gluing conditions (\ref{Bglue}), we infer 
that the D-branes associated with $J = \bar J$ carry a non-vanishing 
2-form potential (B-field) 
\begin{equation}
B \ = \ \frac{1 + \Ad(g)}{1- \Ad(g)} \ .
\label{Bfield} \end{equation}
In the limit $\ik \to \infty$ the three-sphere 
grows and approaches flat 3-space. 
One can parameterize it by a parameter $X$ taking
values in the Lie algebra su(2), such that
$g \approx 1 + X$.
Then, the formula for the $B$-field reads
$$
B \ \approx \ - 2\; \bigl(\ad(X)\bigr)^{-1} \ .
$$
This is the Kirillov 2-form on the spheres in the
algebra su(2)$\, =\, \R^3$.

\noindent
Extrapolating formula (\ref{magic}) to our curved background,
we can construct a bi-vector 
$$
\Theta_A\ =\ \frac{2}{B-B^{-1}}\ =\ 
\frac{1}{2}\; \bigl(\,\Ad(g^{-1}) - \Ad(g)\,\bigr)\ .
$$
Introducing an orthonormal basis $e^a$ in su(2), and
the left- and right-invariant vector fields $e^a_L, e^a_R$ on
the group manifold, one can give an elegant
formula for the bi-vector $\Theta_A$,
$$
\Theta_A\ =\ \frac{1}{2}\; e^a_L \wedge e^a_R \ .
$$
The Schouten bracket of $\Theta_A$ (which generally 
characterizes the deviation from the Jacobi identity)
is of the form
$$
\phi\ :=\ [\,\Theta_A, \Theta_A\,]\ =\ \frac{1}{6}\; f_{abc}\; 
(e^a_L- e^a_R)(e^b_L - e^a_R) (e^c_L - e^c_R)\ .
$$
Here $f_{abc}$ are the Lie algebra structure constants,
the same as those in the expression for the field strength $H$.
This calculation makes sense for an arbitrary simple Lie
group. In general, the right hand side does not vanish 
and gives the obstruction for the Jacobi identity.
In the case of $G\,=\,$SU(2), $\phi$ vanishes for dimensional
reasons: It is a 3-vector tangent to the 
2-dimensional conjugacy classes. 
In the infinite volume limit $\ik \to \infty$, the bi-vector
$\Theta_A$ becomes 
$$
\Theta_A = \ad(X) \ ,
$$
which is the Kirillov-Kostant Poisson bi-vector. 
Consequently, the geometry of the 
limiting theory $\ik = \infty$ is very close to the well-known situation 
of flat branes in a flat background with constant $B$-field, and we 
expect that the world-volume algebras of our branes in the WZW model 
will be quantizations of two-spheres. 
\nno
For finite $\ik$, however, the background is curved and carries a 
non-vanishing NS 3-form $H$. This will result in a non-associative 
deformation of the $\ik= \infty$ theory. Since the three indices of 
the new object $H$ can relate three-fold products with different 
positions of brackets, the violation of associativity will turn out 
to be rather mild. 
\noindent 
The semi-classical extension of the above analysis shows that, for 
fixed gluing conditions, only a finite number of SU(2) conjugacy 
classes satisfy a Dirac-type flux quantization condition \cite{AlSc}. 
These ``integer'' conjugacy classes are the two points $e$ and $-e$ 
along with $\ik - 1$ of the spherical conjugacy classes (those passing 
through the points ${\rm diag}(\exp(i\pi j/\ik), \exp(-i\pi j/\ik))$ 
for $j=1,\ldots,\ik-1$).

\subsection{Exact CFT description.} 
\def\cC{{\cal C}}
The WZW model on the upper half-plane is known in enough detail
to support and specify the rather crude arguments of the previous 
subsection by an exact CFT analysis. In fact, for the situation we 
are dealing with (gluing conditions $J = \bar J$ in a ``parent'' 
CFT on the full complex plane with diagonal modular invariant 
partition function), Cardy \cite{Car} was able to list all \cite{PSS} 
possible boundary conditions. 
There exist $\ik +1$ of them, differing in the bulk field one-point 
functions (brane charges) and labeled by an index $\a = 0, \frac{1}{2}, 
\dots, \frac{\ik}{2}$. Without entering a detailed description of these 
boundary theories \cite{Car}, we recall that their state spaces have the 
form 
\def\cH{{\cal H}}
\be \label{partdec}
 \cH_\a \ = \ {\bigoplus}_J \ N_{\a\a}^J \ \cH^J 
\ee
where $\cH^J$, $J= 0,\frac{1}{2}, \dots, \frac{\ik}{2}$, denote 
irreducible highest weight representations of the affine Lie algebra 
${\widehat{{\rm SU}}(2)}_\ik$, and where $N_{IJ}^K$ are the associated 
fusion rules. Note that only integer spins $J$ appear on the right hand 
side of (\ref{partdec}). 

\smallskip\noindent
There exists a variant of the state-field correspondence which 
assigns a boundary field $\psi(x)$ to each element 
$|\psi\rangle \in \cH_\a$ (see e.g.\ \cite{ReSh1}). In particular, 
the SU(2) WZW boundary theory labeled by $\a$ contains SU(2)-multiplets 
associated to primary boundary fields, namely  
$$ \Psi^J(x) \ = \ (\psi^J_m (x)) \ \ \ \mbox{ with } \ \ \ 
   J = 0,1,\dots, {\min}(2\a, \ik-2\a)  $$ 
and $m = -J,\dots , J$. All these boundary fields are defined 
for arguments $x$ on the real line and their correlators have, 
in general, no unique analytic continuation into the upper half-plane.

\smallskip\noindent
In the flat target case, we chose U(1)-primaries as generating elements 
of the world-volume algebra. Now, it is more appropriate not to break 
the group symmetry by hand and, therefore, to keep the full SU(2)-multiplets
$\Psi^J(x)$. 
For a fixed order $x > y$ of arguments on the real line, the 
OPE of two such boundary fields reads
\be \label{boundOPE}
    \psi^I_i(x)\ \psi^J_j(y) \ \sim \ {\sum}_{K,k}\ (x-y)^{h_I+h_J - 
    h_K} \ \CG{I}{J}{K}{i}{j}{k}\ c^{\ik,\a}_{IJK} \ 
    \psi^K_k(y)\ \ ,    
\ee
where $h_J$ is the conformal dimension of $\Psi^J$ and $[:::]$ denote 
the Clebsch-Gordan coefficients of the group SU(2). The latter simply 
compensate for the different transformation behavior of the fields on 
the left and right hand side under the action of the zero-mode 
subalgebra of ${\widehat{\rm SU}(2)}_\ik$. Hence, the non-trivial 
information in (\ref{boundOPE}) is contained in the new structure 
constants $\cC = (c^{\ik,\alpha}_{IJK})$. 
\nno
In a consistent theory, these must obey sewing constraints, which 
were first analyzed  by Lewellen in \cite{Lew}; see also \cite{PSS}. 
Recently, these constraints were reconsidered by Runkel \cite{Run} 
for the A-series of Virasoro minimal models. His findings carry over to 
SU(2) WZW models on the upper half-plane and show that the only possible 
solution to the sewing constraints is given by the fusing matrix $F$ of 
the WZW theory, 
\be \label{cval1}
c^{\ik,\a}_{IJK} \ = \ \Fus{\a}{K}{\a}{\a}{I}{J}_\ik \ \ . 
\ee   
It is one of the fundamental results on the relation between quantum 
groups and conformal field theory (see e.g.\ \cite{AGGS}) that the 
fusing matrix of the WZW model is obtained from the $6J$ symbols 
of the quantum group algebra $U_q({\rm su}(2))$ according to  
\be \label{cval2} 
\Fus{\a}{K}{\a}{\a}{I}{J}_\ik \ = \ \SJS{I}{J}{K}{\a}{\a}{\a}_q\ \ 
 \ \mbox{ where } \ \ \ q = e^{\frac{2\pi i}{\ik+2}}\ \ . 
\ee
In the limit $q \rightarrow 1$, the $6J$ symbols of the quantum 
group algebra approach those of the classical algebra $U({\rm su}(2))$,   
thus the structure constants $c^{\ik,\a}_{IJK}$ of the boundary OPE 
become $6J$ symbols of the group SU(2) when the level $\ik$ is sent 
to infinity. Note that in this limit, the conformal dimensions 
$h_J = J(J+1)/(k+2)$ tend to zero so that the OPEs (\ref{boundOPE}) 
of boundary fields become regular as in a topological theory. 
\medskip

\section{D-brane geometry, fuzzy two-spheres, and quantum groups} 

We are now prepared to follow the procedure sketched at the end 
of Section 2 and to read off the world-volume geometry of branes 
in the SU(2)-WZW model. So let us think of the boundary fields
$\psi^I_i= V(Y^I_i)$ as being assigned to elements $Y^I_i$ of 
some vector space, and let us use the operator product expansion 
(\ref{boundOPE},\ref{cval1},\ref{cval2}) to define a multiplication 
by the prescription
\be \label{FSOPEk}
 Y^I_i\, \star \, Y^J_j \ = \ {\sum}_{K,k} 
 \;\ \CG{I}{J}{K}{i}{j}{k}\ c^{\ik,\a}_{IJK} \ Y^K_k \ \ .   
\ee
As in (\ref{boundOPE}), the summation on the right hand side runs from 
$K=0$ to a maximal spin $K_{\rm max}= \min(I+J,\ik-I-J,2\a,\ik-2\a)$. 
First, we shall investigate this product in the limiting case $\ik = 
\infty$, where it produces a familiar algebraic structure. 
Passing to finite levels leads to the following two changes: There 
is a $\ik$-dependent deformation of structure constants $\cC$, cf.\ 
(\ref{cval2}), and the range of the summation in (\ref{FSOPEk}) 
becomes a function of the level, $K_{\rm max}= K_{\rm max}(\ik)$. 
We shall separate these two phenomena by looking at an intermediate 
case where $\ik$ is non-rational and where we omit the $\ik$-dependent 
restriction on the $K$-summation. 

\smallskip
\noindent
{\it Infinite level $\ik = \infty$}:\quad Recall that, in the case 
of infinite level, the structure constants $\cC$ in eq.\ (\ref{FSOPEk}) 
are given by the $6J$ symbols of the group SU(2). The semi-classical 
analysis showed that $H \to 0$, so we expect the world-volume algebra 
to be associative. Indeed this can be confirmed using the Biedenharn-Elliot 
(or pentagon) relation for the $6J$ symbols, along with the fact that $6J$ 
symbols of the form (\ref{cval2}) vanish whenever $K > 2\alpha $. Hence, 
for infinite level our relations define an infinite set of associative 
algebras $S^2_\alpha,\  \alpha = 0,\frac12, \dots,$ with finite linear 
bases consisting of \ dim$\,(S^2_\alpha) = (2\alpha + 1)^2$ elements. 
\nno Since 
the dimension of each of these algebras is a perfect square, one may 
already suspect that they are full matrix algebras, i.e.\ that $S^2_\alpha
\cong M_N(\C)$ with $N = 2\alpha + 1$. 
To describe the isomorphism, we first note that $M_N(\C)$ admits an 
action of the group SU(2) by conjugation with group elements evaluated 
in the $N$-dimensional representation of SU(2). Under this action, the 
SU(2)-module $M_N(\C)$ decomposes into a direct sum of irreducible 
representations $V^J$,  
\be \label{FSdec}
M_N(\C)  \ \cong \ {\bigoplus}_{J=0}^{N-1}  \ V^J \ \ . 
\ee
Only integer $J$ appear, so this agrees with the decomposition of 
the state space $\cH_\a,\ \a = (N-1)/2,$ in eq.\ (\ref{partdec}) for 
boundary WZW models at sufficiently large (or infinite) level $\ik$. 
Thus, we can identify our elements $Y^J_j$ with a basis of the spaces 
$V^J$. The isomorphism (\ref{FSdec}) allows to work out multiplication 
rules for any two such basis elements from the multiplication of 
$N\times N$-matrices. The result \cite{Hop} turns out to coincide with 
our formula (\ref{FSOPEk}), which shows that $S^2_\alpha$ and $M_N(\C),\ 
N = 2\alpha + 1,$ are indeed isomorphic as associative algebras. 
\nno
The non-commutative spaces $S^2_\alpha$ are known as {\it fuzzy spheres} 
and are obtained when one quantizes functions on a two-sphere with the 
usual Poisson structure (see e.g. \cite{Mad} and references therein). 
The two-spheres may also be identified with co-adjoint orbits of SU(2). 
According to Kirillov, their quantization gives all representations of 
the Lie algebra su(2) or of its universal enveloping algebra $U({\rm 
su(2)})$. Note that the size $N = 2\a + 1$ of our matrices agrees with 
the number of components for an su(2)-multiplet of spin $\a$. Hence, 
through the investigation of maximally symmetric branes on SU(2) at 
$\ik = \infty$, we have recovered Kirillov's theory of co-adjoint orbits. 
\medskip

\noindent 
{\it Finite non-rational level $\ik$}:\quad Let us stress that this 
case does not appear among the exact boundary theories above (for 
non-compact WZW models, it is the generic situation). We include it  
here merely as an intermediate step before presenting the structure for 
finite integer level $\ik$. To be more precise, we 
consider the algebras spanned by $Y^J_j$ with relations (\ref{FSOPEk}) 
in which the structure constants $\cC$ are given by the $6J$ symbols 
(\ref{cval2}) of the quantum group algebra $U_q({\rm su(2)})$, 
but with summation over the same range as in the case $\ik = \infty$. 
\nno
The resulting algebras $S^2_{\a,q}$ with $q = \exp(2\pi i/(\ik+2))$ not 
a root of unity cease to be associative. But they are still 
quasi-associative in the sense that 
\be Y^I_i \, \star \, (Y^J_j \,  \star\, Y^K_k)  
   (\tau^I_{in} \o \tau^J_{jm} \o \tau^K_{kl})(\varphi)\ = \  
   \   (Y^I_n \, \star \, Y^J_m )\,  \star\, Y^K_l \label{qassoc} \ee
where the $\tau^L$ denote representations of $U({\rm su(2)})$ and where 
$\varphi \in U({\rm su(2)})^{\otimes 3}$ is Drinfeld's ``re-associator'' 
\cite{Dri}. The proof of this statement is sketched in the appendix. 

\noindent 
When we perform a standard quasi-classical limit, commutators are  
replaced by the  brackets corresponding to the bi-vector
$\Theta_A$. For a general compact simple Lie group $\Theta_A$
fails to satisfy the Jacobi identity. This corresponds to
the leading non-vanishing term in the 
$\frac{1}{\ik}$-expansion of the re-associator $\varphi$,
$$
\varphi = 1 + \frac{1}{6 \ik} f_{abc} e^a \otimes e^b \otimes e^c + \dots
$$
where $e^a$ is, as above, an orthonormal basis in the Lie algebra,
and $f_{abc}$ are the corresponding structure constants.
When applied to the relation (\ref{qassoc}), the Lie algebra
generators $e^a$ act by the adjoint vector fields $(e^a_L - e^a_R)$.
In the case of $G\,=\,$SU(2) this leads to vanishing of the 
first order correction to the associativity law. This is
in accordance with vanishing of $[\Theta_A, \Theta_A]$ in this
case. Note that even in the SU(2) case higher
order corrections to the associativity law do not vanish.

\noindent 
Let us briefly mention that our quasi-associative algebras 
$S^2_{\a,q}$ are closely connected to associative deformations
of the fuzzy sphere which employ the Clebsch-Gordan coefficients 
of the deformed $U_q({\rm su(2)})$ instead of their classical 
analogs. Some details on these algebras and their associativity 
can be found in the appendix. For now, let us only remark that they 
are factors of the quantum spheres introduced by Podle\'s in 
\cite{Pod}. Their relation to our algebras $S^2_{\a,q}$ is
based on the fact that one can obtain the Clebsch-Gordon maps of 
classical Lie algebras from their $q$-deformed counterparts 
with the help of Drinfeld's ``twist element'' 
$F \in  U({\rm su(2)})^{\otimes 2}$. The latter provides the 
following factorization formula for the re-associator: 
$$ \varphi \ = \ (\id \o \Delta)( F^{-1})\, (e \o F^{-1}) 
      \, (F \o e)\,  (\Delta \o \id)( F ) \ \ $$
where $\Delta$ denotes the co-product of $U({\rm su(2)})$. 
Combining these two roles of the twist element $F$, one can 
show that our algebras $S^2_{\a,q}$ are ``twist equivalent'' 
to  associative factors of a Podle\'s sphere or, more 
explicitly, to the same matrix algebras $M_N(\C), \ N = 2 
\alpha + 1,$ as in the case of infinite level. Hence, we 
simply recover the representations for the usual 
$q$-deformation of $U({\rm su(2)})$ at generic values of 
the deformation parameter. 
\medskip
   
\noindent
{\it Finite integer level $\ik$}:\quad The associated  algebras 
$A^{\ik}_\a$ are spanned by the generators  $Y^J_m$ with the label $J$ 
chosen from the set $J = 0,1,\dots, \min(2 \a, \ik-2\a)$. Multiplication 
of these elements is defined through eq.\ (\ref{FSOPEk}) with structure 
constants $\cC$ now given by the $6J$ symbols of $U_q({\rm su(2)})$ 
at the root of unity $q = \exp(2\pi i/(\ik +2))$. In addition, the 
summation on the right hand side is now restricted to run from $K=0$ 
to $\min(I+J,\ik-I-J,2\a, \ik-2\a)$. Viewed as SU(2)-modules, the 
linear spaces $A^{\ik}_\a$ decompose as follows:  
$$A^\ik_\a \ \cong \ \left\{  \begin{array}{ll} 
   S^2_\a & \ \mbox{ for } \ \ 0 \leq  \a \leq \frac{\ik}{4}\\[2mm] 
   S^2_{\ik/2 - \a} & \ \mbox{ for } \ \ \frac{\ik}{4} \leq \a \leq 
                         \frac{\ik}{2}  \end{array} \right. \ \ . 
$$
Again, the algebras $A^{\ik}_\a$ are only quasi-associative, and they 
provide examples of the geometries considered in \cite{MaSh}. Using the 
concept of representations introduced in \cite{Sch}, it is not difficult 
to show that each of the quasi-associative algebras $A^\ik_\a$ possesses
precisely one indecomposable representation on a vector space $W^\a$ of 
dimension
$$\mbox{dim\/}\, W^\a  \ = \ \left\{  \begin{array}{ll} 
   2\a+1 & \ \mbox{ for } \ \ 0 \leq  \a \leq \frac{\ik}{4}\\[2mm] 
   \ik - 2\a + 1 & \ \mbox{ for } \ \ \frac{\ik}{4} \leq \a \leq 
                         \frac{\ik}{2}  \end{array} \right. \ \ . 
$$  
According to our previous discussion, the algebras $A^\ik_\a$ and their 
representations on $W^\a,\ \a = 0,\frac12, \dots, \frac{\ik}{2}$, generalize 
Kirillov's theory of co-adjoint orbits to quantum groups at roots of unity.
In other words, the algebras $A^\ik_\a$ we obtain are ``quantizations''
of integer conjugacy classes on SU(2). Summing over all possible brane 
sectors, i.e.\ over the index $\alpha$, we construct a deformed universal 
enveloping algebra. 

\noindent
Of course, quantum group algebras were constructed within the framework 
of chiral conformal field theory before, see e.g.\ \cite{FrK,MoRe,AlSh,Dri}. 
As long as we avoid roots of unity, our new derivation from boundary 
conformal field theory reproduces well-known algebraic structures. 
Differences between the two approaches occur only when $q$ is a root 
of unity. In that case, boundary conformal field theory improves upon 
the old constructions in two respects. First of all, the theory gives 
``physical'' representations exclusively so that there is no need for 
additional truncations. Furthermore, the dimensions $\dim W^\a$ of the 
representation spaces are invariant under the simple current symmetry 
which interchanges $\a$ and $\ik/2-\a$.    

\noindent
When we increase the level $\ik$, the radius of the three-sphere 
grows and we can fit more and more branes into the background. At 
the same time, the 3-form field strength decreases and the world-volume 
algebras become ``more associative'' --  while their non-commutativity 
survives. 
\nno
This is to be compared to the non-commutative targets obtained in 
\cite{FG,FGR,Gr} from closed strings: The $\ik \to \infty$ limit of 
these targets is simply the classical group SU(2). The different 
behavior of closed and open string geometry may be explained 
as follows: Both closed and open strings feel the presence of the 
NS 3-form field $H$ at finite level. Open strings are also sensitive to 
the concrete choice of a 2-form potential $B$, while closed strings 
``see'' only its cohomology class. In the flat space limit 
$\ik = \infty$, the cohomology becomes trivial while $B$ itself 
stays non-zero and is responsible for non-commutativity on the brane. 

\section{Summary and outlook}
We have derived non-commutative world-volume algebras for 
D-branes in the SU(2) WZW model, using a general scheme that 
can be applied to arbitrary branes given as conformal boundary 
conditions, including supersymmetric cases. In the process, we 
have seen how abstract objects 
from the CFT description, like Cardy's boundary states and 
Runkel's OPE coefficients, acquire a geometrical meaning -- if in 
terms of non-commutative (and sometimes non-associative) spaces. 
The SU(2) WZW model provides just the simplest example of 
a string background with a non-vanishing 3-form field 
strength $H$, but we think that it illustrates quite nicely 
much of the behavior one should expect from more complicated 
backgrounds. In particular, the discussion of SU(2) branes carries 
over to boundary WZW models with other structure groups G (at least 
in the compact case) and leads 
to a quantization of integer conjugacy classes in G. It might 
be interesting to investigate also branes that are not maximally 
symmetric, i.e.\ where the gluing conditions respect only a 
subalgebra of the maximal chiral symmetry algebra \cite{JFCS}. 

\noindent 
Boundary CFT yields world-volumes independently of whether  
limiting classical pictures are available or not, and it  
actually provides more structure than a mere set of 
non-commutative algebras. Connes' program \cite{Con} shows that, in 
order to talk about the geometry of a non-commutative space, it is 
necessary to fix further ``spectral data'', including a Hilbert 
space on which the (associative) world-volume algebra and a 
generalized Dirac or Laplace operator act. 
How these data can be extracted from a CFT has been discussed, for 
the bulk case, in \cite{FG,FGR}. The importance of the Laplace 
operator, which is related to the conformal Hamiltonian 
$L_0$, can also be seen in the context of our definition of 
non-commutative world-volumes: In order to re-derive the OPE of 
boundary operators from the algebraic structure of the world-volume, the
spectrum of conformal dimensions must be known, cf.\ the remark after 
eq.\ (\ref{pointw}). 
\nno
In a CFT on the upper half-plane, additional structure is available, 
e.g.\ in the form of boundary condition changing operators which induce 
transitions between two different boundary conditions $\a,\b$. The OPE 
of the boundary fields $\Psi^I(x)$ with boundary condition changing 
operators gives rise to bi-modules $B_{\a\b}$ over the world-volume 
algebras of the two associated branes. In the case of D-branes on a 
group manifold, these bi-modules allow to construct tensor products 
for representations of the associated quantum group. OPEs involving 
two boundary condition changing operators provide even more data, 
namely a full braided tensor category.

\smallskip\noindent
Some comments on our general scheme to extract a world-volume 
algebras from the boundary CFT description of branes are in order. 
It involves a choice of ``generating elements'' among the boundary 
fields. From a pure CFT perspective, one could restrict to primary 
operators only, or one could work with all boundary operators and 
thus with an infinite-dimensional world-volume. In a sense, the 
latter algebra would include all internal excitations of the 
``static'' space defined using primary fields. The WZW case, where 
it proved natural to keep the full group multiplets associated with 
primary boundary fields, suggests that there are distinguished 
``intermediate'' choices. For a large class of CFTs, the appropriate 
generalization of the lowest-dimension spaces of WZW models is likely 
to be given by the special subspaces introduced in \cite{Na}; see 
also \cite{ARS}. 

\smallskip\noindent
Placing the CFT into a string theory context can remove the 
arbitrariness and provide clear guidelines as to which world-volume 
generators to select from the boundary fields: String theory contains 
additional parameters like $\alpha'$, and the relevant generators of 
the world-volume algebra are those surviving in some limiting regime. 
E.g.\ in the flat background case, one can remove all higher 
excitations by sending $\alpha'$ to zero while keeping the $B$-field 
finite; see \cite{SeWi} and also \cite{DoHu}. 
It may be possible that a number of interesting limits exists; then one 
expects that the world-volume of a brane can look very different 
in different regimes, and that full string theory can ``interpolate'' 
between those geometries.

\noindent 
The next task would be to calculate the effective action on the -- 
in general non-com\-mu\-tative --  world-volume of the brane. The 
lowest-order terms are, of course, already given by our ``multiplication 
table'' (the OPE coefficients). In principle, higher-order contributions 
can be computed from the same data, but in practice one still needs to 
integrate over world-sheet moduli. 
\nno
In the context of the Douglas-Hull model, the effective field theories 
were found to be non-commutative supersymmetric gauge theories with some 
amount of non-locality \cite{CDS,DoHu,effFT,NCTor,ChHo,ArVo}. Seiberg 
and Witten could show that these models are equivalent to ordinary gauge 
theories on a flat brane \cite{SeWi}. It remains to be seen whether 
classical structures are stretched further when more general CFT 
backgrounds are taken as a starting point. Perhaps it is worthwhile 
to compare the induced field theories with existing models on fuzzy 
geometries (see e.g.\ \cite{FFT}). 

\noindent 
It would also be interesting to investigate further the relation 
between world-volume non-commutativity as introduced in \cite{DoHu} 
and non-commuting moduli as discovered by Witten \cite{WiBst}. 
Both phenomena can be traced back to failures in locality 
properties of boundary fields -- see \cite{RS2,RS3} for the case 
of moduli -- so that there exists a direct connection between the 
brane's intrinsic ``fuzziness'' and the way it ``perceives'' its 
ambient target. \\[10mm]
\noindent 
{\bf Acknowledgements:} 
We would like to thank I.\ Brunner, 
C.\ Chu, R.\ Dijkgraaf, M.\ Douglas, J.\ Fr\"ohlich, J.\ Fuchs 
K.\ \gaw, O.\ Grandjean, P.\ Ho, J.\ Hoppe, C.\ Klim\v c\'\i k, 
N.\ Landsman, 
G.\ Moore, A.\ Polychronakos, G.\ Reiter, A.\ Schwarz, C.\ Schweigert, 
S.\ Shatashvili, I.T.\ Todorov and especially J.\ Teschner for 
useful and stimulating discussions. 
V.S.\ is grateful to the DAAD for support and to the AEI Potsdam 
for hospitality. 

\bigskip\noindent
{\bf Note added:} After this work was completed, another approach 
to the geometry of branes in WZW models based on exact CFT methods 
was presented in \cite{FFFS}. 

\section*{Appendix: (Quasi-)associativity} 
\phantom{a}  

\renewcommand{\theequation}{{A.\arabic{equation}}}
Here we collect some basic material on 
Clebsch-Gordan maps, $6J$-symbols and the (quasi-)associativity 
of various algebras mentioned in the main text. Let us denote 
by $\tau^I$ the irreducible representation of $U_q({\rm su(2)})$ 
with spin $I$. By definition, Clebsch-Gordan maps $C_q(IJ|K): 
V^I \o V^J \rightarrow V^K$ intertwine between the actions of 
$U_q({\rm su(2)})$ on the product module $V^I \o V^J$ and the 
irreducible module $V^K$. 
$6J$ symbols enter the theory through the basic relation  
\be C_q(MK|L) \, (C_q(IJ|M) \o \id^K) \ = \ \sum_P 
    \SJS{L}{K}{M}{I}{J}{P}_q \,C_q(IP|L)\, (\id^I \o C_q(JK|P)) 
\ \ .\label{eqA1}\ee
They obey a number of fundamental equations. For our purposes, 
the Biedenharn-Elliot (pentagon) relation is the most important 
one. With the spin labels set to the values that we need 
below, it implies
\be \sum_M \SJS{L}{K}{M}{I}{J}{P}_q  \,
      \SJS{I}{J}{M}{\alpha}{\alpha}{\alpha}_q  \,    
        \SJS{M}{K}{L}{\alpha}{\alpha}{\alpha}_q
 \ = \  \SJS{J}{K}{P}{\alpha}{\alpha}{\alpha}_q  \,    
        \SJS{I}{P}{L}{\alpha}{\alpha}{\alpha}_q
\label{eqA2} \ee  
Relations (\ref{eqA1},\ref{eqA2}) hold for generic $q$ and at the 
classical point $q=1$ where we are dealing with representation 
theory of ordinary Lie algebras. 

\medskip
\noindent 
Let us now study the algebra generated by $Y^I_i$ for $I = 0,1, 
\dots 2\alpha$ and $|i| \leq I$ with the multiplication rules
\be Y^I_i \star Y^J_j \ = \ \sum_{K,k}\  \CG{I}{J}{K}{i}{j}{k}_q \,
   \SJS{I}{J}{K}{\alpha}{\alpha}{\alpha}_q\ Y^K_k \ \ . 
\label{qfuzzy}\ee 
The Clebsch-Gordan coefficients on the right hand side are obtained from 
the maps $C(IJ|K)$ once we have selected a basis in each representation 
space $V^L$. Associativity of this algebra is rather easy to prove with 
the help of eqs.\ (\ref{eqA1}) and (\ref{eqA2}): 
\ba 
(\, Y^I_i \star Y^J_j)\, \star Y^K_k & = & 
 \sum_{L,l,M,m} \hspace{15pt} 
  \CG{I}{J}{M}{i}{j}{m}_q \,\CG{M}{K}{L}{m}{k}{l}_q\, 
     \SJS{I}{J}{M}{\alpha}{\alpha}{\alpha}_q \,     
        \SJS{M}{K}{L}{\alpha}{\alpha}{\alpha}_q \ Y^L_l \nn \\[2mm] 
 & = & 
    \sum_{L,l,M,P,p} \ \ \CG{J}{K}{P}{j}{k}{p}_q\, \CG{I}{P}{L}{i}{p}{l}_q 
       \SJS{L}{K}{M}{I}{J}{P}_q \,
      \SJS{I}{J}{M}{\alpha}{\alpha}{\alpha}_q \,     
        \SJS{M}{K}{L}{\alpha}{\alpha}{\alpha}_q\  Y^L_l \nn \\[2mm]
 & = & 
    \sum_{L,l,P,p}\ \hspace{15pt} \CG{J}{K}{P}{j}{k}{p}_q \,
    \CG{I}{P}{L}{i}{p}{l}_q  \, \SJS{J}{K}{P}{\alpha}{\alpha}{\alpha}_q      
     \,   \SJS{I}{P}{L}{\alpha}{\alpha}{\alpha}_q \ Y^L_l \nn \\[2mm] 
 & = & Y^I_i \star \, (\, Y^J_j \star Y^K_k) \nn 
\ea 
For the special case $q=1$ this computation proves the 
associativity of the world-volume algebra in the limit $\ik = 
\infty$. When the level $\ik$ is finite and non-rational, however, 
the defining relation for our algebra $S^2_{\a,q}$ from Sect.\ 4 
employs the {\em undeformed} Clebsch-Gordan maps along with 
the deformed $6J$ symbols. Hence, using relation (\ref{eqA1}) 
for $q=1$, we generate an undeformed $6J$ symbol in our 
computation above. The latter cannot be absorbed with the help 
of the pentagon identity, since we have to deal with a product 
of one undeformed and two deformed $6J$ symbols.  

\noindent 
At this point, Drinfeld's re-associator $\varphi \in 
U_q({\rm su(2)})^\o_3$ plays a decisive role because of 
its fundamental property  
\ba 
C(MK|L)\,(C(IJ|M)\o\id^K)(\varphi^{-1})^{IJK}  
&=&\hspace*{-3pt}\sum_P\SJS{L}{K}{M}{I}{J}{P}_q \,C(IP|L)\,(\id^I\o C(JK|P))
   \nn \\[2mm]   
\mbox{where } \ \ (\varphi^{-1}) ^{IJK} \ = \  (\tau^I \o \tau^J \o \tau^K) 
   & & \hspace*{-10mm} (\varphi^{-1}): V^I \o V^J \o V^K \rightarrow  
   V^I \o V^J \o V^K  \ .\nn 
\ea     
Note that this relation involves Clebsch-Gordan maps of the 
Lie algebra and $q$-deformed $6J$-symbols at the same time. $\varphi$ 
allows to modify the proof we have given for the associativity of the 
algebra (\ref{qfuzzy}) such that we obtain the quasi-associativity 
property (\ref{qassoc}). 

\medskip
\noindent  
A relation between our quasi-associative algebra $S^2_{\a,q}$  
and the associative $q$-deform\-a\-tion of the fuzzy sphere can 
be established with the help of Drinfeld's twist element $F$. 
By definition, it maps the deformed and undeformed Clebsch
Gordan maps onto each other,   
$$ C_q(IJ|K) (\tau^I \o \tau^J) (F) \ = \ C(IJ|K) \ \ . $$
This property becomes crucial in showing that the quasi-associative 
algebra for non-rational $\ik$ is ``twist-equivalent'' to the 
associative $q$-deformed fuzzy sphere. Some details on the notion 
of twist equivalence can be found e.g.\ in Section 7.3 of 
\cite{AGS1}.    

\newpage
\newcommand{\sbibitem}[1]{\vspace*{-1.5ex}\bibitem{#1}}

\end{document}